\documentclass[conference]{IEEEtran}
\IEEEoverridecommandlockouts
\usepackage{cite}
\usepackage{amsmath,amssymb,amsfonts}
\usepackage{algorithmic}
\usepackage{graphicx}
\usepackage{textcomp}
\usepackage{xcolor}
\usepackage{multirow}
\usepackage{array}
\usepackage[table]{xcolor}
\ifCLASSOPTIONcompsoc
    \usepackage[caption=false, font=normalsize, labelfont=sf, textfont=sf]{subfig}
\else
\usepackage[caption=false, font=footnotesize]{subfig}
\def\BibTeX{{\rm B\kern-.05em{\sc i\kern-.025em b}\kern-.08em
    T\kern-.1667em\lower.7ex\hbox{E}\kern-.125emX}}
\begin{document}

\title{Gamma2Patterns: Deep Cognitive Attention Region Identification and Gamma–Alpha Pattern Analysis \\

}

\author{
    \IEEEauthorblockN{
        Sobhana Jahan\textsuperscript{1}, 
        Saydul Akbar Murad\textsuperscript{2},
        Nick Rahimi\textsuperscript{2}, 
        Noorbakhsh Amiri Golilarz\textsuperscript{1}
    }

    \IEEEauthorblockA{
        \textsuperscript{1}Department of Computer Science, The University of Alabama, Tuscaloosa, AL, USA\\
    }
    
    \IEEEauthorblockA{
        \textsuperscript{2}School of Computing Sciences \& Computer Engineering, University of Southern Mississippi, Hattiesburg, MS, USA\\
        Email:  sjahan2@crimson.ua.edu and noor.amiri@ua.edu
    }
}

\maketitle

\begin{abstract}
Deep cognitive attention is characterized by heightened gamma oscillations and coordinated visual behavior. Despite the physiological importance of these mechanisms, computational studies rarely synthesize these modalities or identify the neural regions most responsible for sustained focus. To address this gap, this work introduces Gamma2Patterns, a multimodal framework that characterizes deep cognitive attention by leveraging complementary Gamma- and Alpha-band EEG activity alongside Eye-tracking measurements. Using the SEED-IV dataset \cite{zheng2018emotionmeter}, we extract spectral power, burst-based temporal dynamics, and fixation–saccade–pupil signals across 62 channels or electrodes to analyze how neural activation differs between high-focus (Gamma-dominant) and low-focus (Alpha-dominant) states. Our findings reveal that frontopolar, temporal, anterior frontal, and parieto-occipital regions exhibit the strongest Gamma power and burst rates, indicating their dominant role in deep attentional engagement, while Eye-tracking signals confirm complementary contributions from frontal, frontopolar, and frontotemporal regions. Furthermore, we show that Gamma power and burst duration provide more discriminative markers of deep focus than Alpha power alone, demonstrating their value for attention decoding. Collectively, these results establish a multimodal, evidence-based map of cortical regions and oscillatory signatures underlying deep focus, providing a neurophysiological foundation for future brain-inspired attention mechanisms in AI systems.
\end{abstract}





\section{Introduction}


Deep cognitive attention refers to a form of sustained attention where the relevant information to the task is being processed, and distractions are kept to a minimum \cite{hayles2007hyper}. It requires various neural activities and different oscillations that correspond to different attentional states, and it has been researched using EEG techniques, which are essential to studies related to neuroscience and human-computer interaction \cite{mountcastle1978brain}. Deep attention models and theories are highly important for applications in neuroscience, human-computer interaction, education, cognitive health, and cognitive intelligence-inspired artificial intelligence systems. Deep attention can facilitate more complex cognitive tasks like learning and decision-making.  Attention deficits could cause cognitive exhaustion and different neurological disorders \cite{jonsdottir2013cognitive, kuppuswamy2023role}. Therefore, it is important to provide appropriate biological indices or brain regions of deep attention for more effective attention-tracking systems.


Gamma oscillations (31-50 Hz) \cite{fernandez2023over} are neural signatures of cognitive processing and attention, arising from the organized firing of neurons. There is a notable enhancement of Gamma oscillations in relevant brain regions during deep attention, which facilitates communication among diverse neural assemblies, thereby supporting working memory and perceptual binding. Increases in Gamma power and bursts features reflect cognitive demand and attention \cite{buzsaki2012mechanisms}. On the other hand, Alpha oscillations (8-14 Hz) \cite{jensen2010shaping} are associated with lower brain excitability and are linked to inhibitory control, dominating periods of relaxed and low-cognitive states. Gamma oscillations indicate that there is active processing of information, while Alpha oscillations indicate that they are functionally disengaged from other stimuli. Therefore, the Gamma and Alpha bands play distinct roles in the regulation of attention and cognitive processes.

Eye-tracking features provide additional behavioral data that can enhance EEG analysis of deep cognitive attention. Fixation duration indicates ongoing processing and gaze duration, saccade movement provides insight into fixation point stability, and pupil dilation indicates the level of cognitive load. Together with EEG, especially Gamma-band data, these metrics can contribute to the verification of deep attention engagement. Gamma power and burst features, which correlate with extended fixation points, moderate saccadic movement, and higher pupil dilation; can thus serve as evidence of deep attention engagement \cite{khushaba2013consumer}.

To date, neural oscillations are considered an integral part of attentional regulation and information processing, especially in the Alpha and Gamma bands.
Gamma activity is highly associated with high-level cognitive effort and sustained attention \cite{cohen2014analyzing}. In contrast, Alpha activity reflects inhibitory control and reduced engagement. Eye-tracking features, such as fixation, saccades, and pupil dilation, also provide informative behavioral markers on attentional states \cite{klimesch2007eeg}.
Together, these neural and behavioral modalities offer a rich foundation for studying deep cognitive focus.
However, despite their relevance, the relationship between EEG Alpha–Gamma patterns and Eye-movement behavior during deep focus has not been adequately investigated.

While past studies have often emphasized spectral properties of EEG signals or the behavioral markers from Eye movements, they have predominantly done so in isolation, providing fragmented rather than holistic insights into attentional processes \cite{rayner2009eye}. Consequently, how oscillatory activity, especially from higher frequency bands reflects sustained focus when synchronized with Eye-movement behavior has only begun to be explored \cite{hopfinger2000neural}. Paralleling the shift toward multimodal diagnostics in healthcare, there is significant potential that the integration of neural oscillations with Eye-tracking measures may provide a more complete representation of attentional states \cite{zhu2023review}. Therefore, the neurophysiological basis of sustained focus, particularly regarding localized Gamma activation and burst characteristics, remains unclear. Thus, a critical need exists for a multimodal understanding of these brain regions and oscillatory signals.

Recognizing the necessity of analyzing brain activity, researchers have increasingly employed diverse computational frameworks to classify cognitive states. Deep learning approaches, such as Convolutional Neural Networks (CNNs) \cite{o2015introduction} and Bidirectional Long Short-Term Memory (BiLSTMs) \cite{siami2019performance}, have been utilized to detect high-Gamma activity and temporal patterns with promising accuracy, yet these methods are often limited by an exclusive focus on single frequency bands or susceptibility to subject variability \cite{avital2024cognitive, sridhar2020eeg}. Shifting toward connectivity, other frameworks have investigated functional network features and Phase Locking Values (PLV) to assess emotional states and cognitive loads; while these studies successfully demonstrate that functional connectivity varies with workload, they frequently lack comprehensive temporal dynamics or face challenges regarding ecological validity \cite{yang2020high, panwar2024eeg}. Furthermore, general machine learning models have been developed to classify attention levels using extensive time-frequency features, but despite good generalization, these approaches are often hindered by imbalanced class distributions \cite{mohamed2018characterizing}.

Despite promising progress, review of existing literature reveals that most prior studies have focused primarily on EEG-based classification \cite{craik2019deep}, such as emotion recognition or frequency band discrimination, rather than understanding the underlying neural mechanisms of attention. Many works also suffer from notable limitations, including imbalanced datasets, limited robustness, limited to laboratory conditions \cite{ismail2020applications}, and the absence of multimodal assessment \cite{vanneste2021towards}. Moreover, studies that attempt to visualize regional activity often report subject-specific topography maps for only a few participants, without providing a generalized representation of the cortical regions involved \cite{avital2024cognitive}. These studies also tend to lack clear identification or interpretation of the brain areas most relevant to cognitive attention. As a result, there remains a meaningful gap in the literature: the field lacks a comprehensive, multimodal framework capable of determining which brain regions dominate during deep cognitive focus and how gamma-dominant patterns differ from Alpha-dominant activity. Addressing this gap provides the motivation for the present study. 

To overcome these limitations, we introduce Gamma2Patterns, the first  multimodal framework that leverages EEG spectral power and burst characteristics alongside Eye-tracking measures to characterize attention-related neural patterns. The approach identifies influential regions of the brain, quantifies differences between Alpha and Gamma Bands, and extracts Gamma power and burst statistics relevant for deep-focus and low-focus states.
The key contributions of this study are:
\begin{itemize}

\item  The study identifies the key cortical regions associated with deep-focus cognition through analysis of discriminative Gamma-band spatial patterns.
\item  The work differentiates the signal characteristics of Alpha and Gamma bands to highlight their contrasting roles in cognitive focus.
\item  The approach extracts Gamma power and burst-duration features from EEG recordings to enhance deep-focus state characterization.

\end{itemize}


The rest of the paper is organized as follows: Section II discusses the methodology. Here, the entire work is divided by extracting the Alpha and Gamma bands, calculating the power, classification, explainability, power intensity analysis, and topographical map.  Section III presents the Experimental Results and Discussion. Dataset description, all the classification results, LIME results, and power-based analysis are mentioned along with a clear discussion in this section. Section IV consists of concluding remarks.



\section{Methodology}

\begin{figure*}
    \centering
    \includegraphics[width=0.9\linewidth]{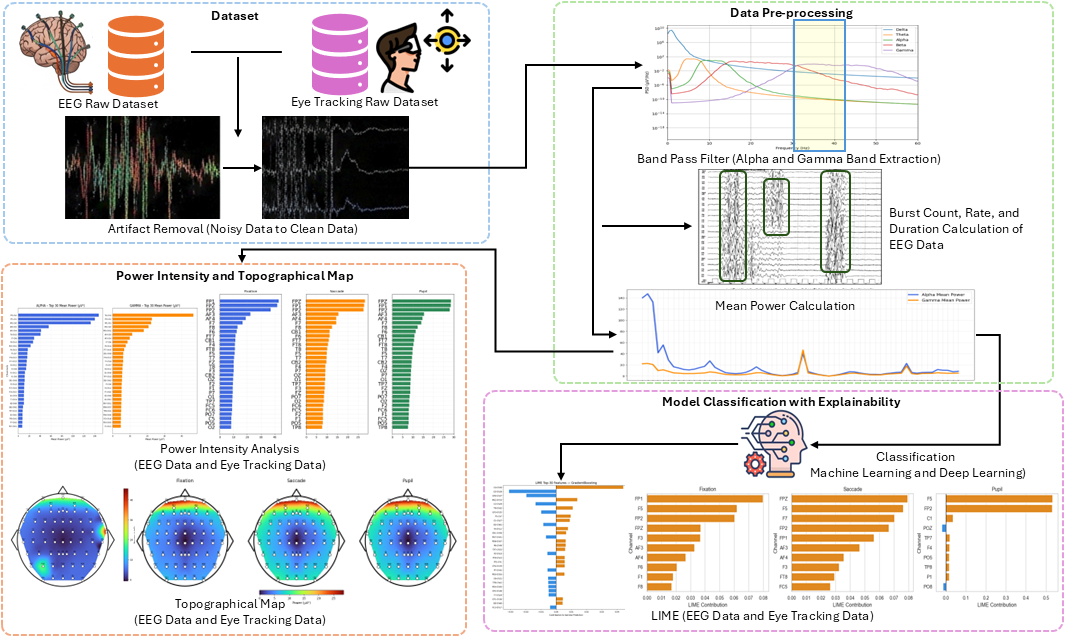}
    \caption{Overview of the proposed model architecture for identifying brain regions associated with deep cognitive attention. Raw signals undergo artifact removal and preprocessing, including Alpha–Gamma band extraction, power computation, and burst analysis. The framework then performs classification with LIME-based explainability to assess feature relevance. In addition, power intensity visualizations and topographical brain maps are generated to highlight channel-wise feature importance and localize cortical regions associated with deep cognitive focus.}
    \label{methodology}
\end{figure*}

The methodology is organized into three primary phases: Data Pre-processing, Model Classification with Explainability, and Power Intensity \& Topographical Map, with the specific details regarding dataset acquisition discussed separately in Section III. As illustrated in Fig. \ref{methodology}, the pipeline begins with Data Pre-processing, where raw EEG and Eye-tracking signals undergo artifact removal, band-pass filtering, and feature extraction-specifically calculating mean power and burst metrics (count, rate, and duration). This is followed by Model Classification with Explainability, which employs machine learning and deep learning classifiers to distinguish cognitive states, utilizing LIME to interpret feature importance. Finally, the framework produces a Power Intensity \& Topographical Map, visualizing channel-wise activity to localize the specific brain regions driving deep cognitive attention.

\subsection{Data Pre-Processing}

In this research we have used SEED-IV \cite{zheng2018emotionmeter} dataset. The SEED-IV dataset contains the EEG and Eye tracking data of 45 participants for all five bands (Delta, Theta, Alpha, Beta, and Gamma). The details about the dataset is mentioned in the Section III Subsection A. As we have EEG data and Eye tracking Data, we have pre-processed these data separately. The detailed description of the pre-processing procedure is mentioned below.

\subsubsection{EEG Data}
Five frequency bands were extracted from the raw EEG signals (.mat) across all 62 channels. For band extraction, a Butterworth band-pass filter was employed to isolate the frequency components of interest. This filter was selected for its smooth frequency response, which enables effective separation of the desired bands while attenuating out-of-band frequencies. In this study, Alpha and Gamma-band activities were isolated by suppressing all frequency components outside the 8–14 Hz and 31–50 Hz ranges, respectively. Each EEG channel was treated as an independent time series, allowing band-specific signals to be processed separately. The filter design normalized the upper and lower cutoff frequencies relative to the Nyquist frequency, which was set to 500 Hz (half of the 1000 Hz sampling rate). To prevent phase distortion, a zero-phase forward–backward filtering technique was applied.

\begin{figure}[htbp]
\centerline{\includegraphics[width=0.5\textwidth]{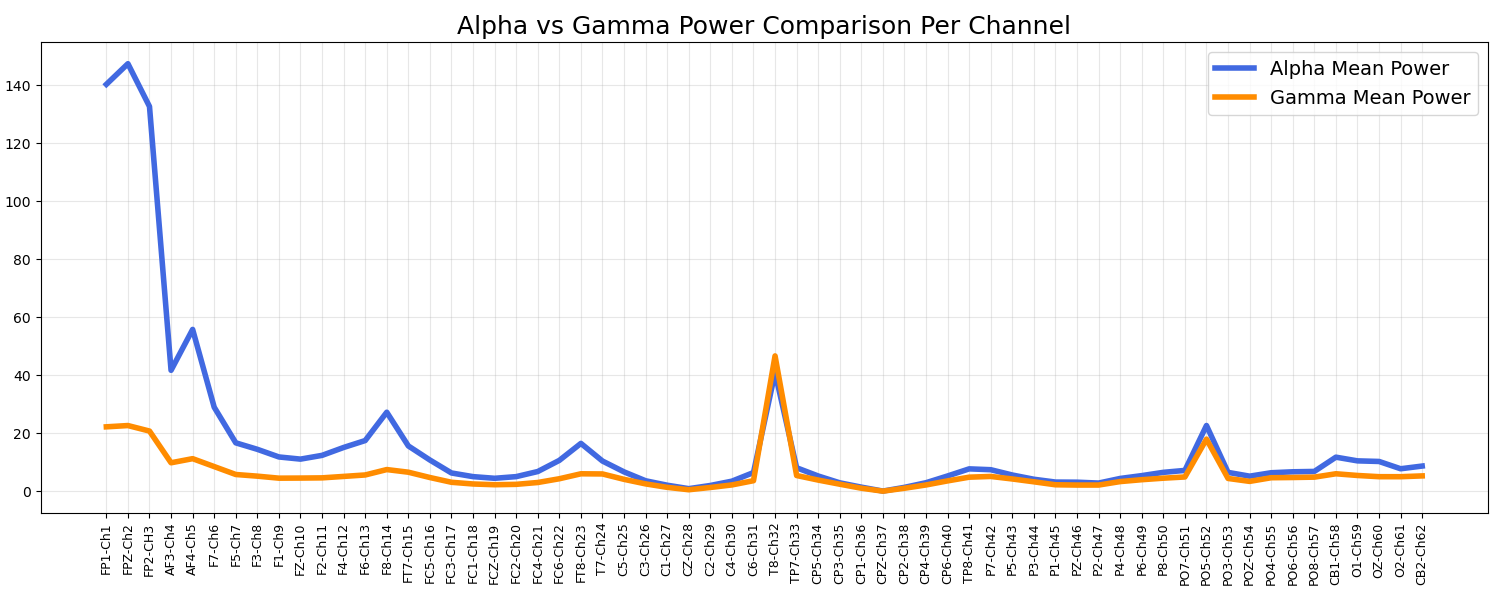}}
\caption{Alpha vs Gamma power comparison per channel.}
\label{alpha_gamma_power}
\end{figure}

Following filtration, the signal strength or energy of the oscillatory waves was quantified by calculating the power for each channel. EEG electrodes measure electrical potential differences (voltage) over time. Because the effective resistance of the recording circuit is constant for a given channel, the instantaneous power $p(t)$ is proportional to the square of the voltage $v(t)$. This relationship is derived from Ohm's Law \cite{franco1995electric}:
\begin{equation}
p(t) = v(t)\, i(t) 
\label{eq:power_def}
\end{equation}
where $v(t)$ is the voltage and $i(t)$ is the current. Substituting the current $i(t) = v(t)/R$, the equation becomes:
\begin{equation}
p(t) = v(t) \left( \frac{v(t)}{R} \right) = \frac{v(t)^2}{R}
\label{eq:power_sub}
\end{equation}
Given a fixed resistive load $R$, the relationship simplifies to:
\begin{equation}
p(t) \propto v(t)^2
\label{eq:proportionality}
\end{equation}
Consequently, the squared voltage serves as the fundamental quantity for estimating power. The mean power for a given segment is calculated as the average of these squared voltage values over $N$ time samples \cite{abdul2004power}:
\begin{equation}
\text{Power} = \frac{1}{N} \sum_{t=1}^{N} v(t)^2
\label{eq:mean_power}
\end{equation}
where $v(t)$ denotes the EEG voltage at time point $t$ (measured in microvolts), and $N$ represents the total number of samples in the segment. The resulting unit of power is $\mu V^2$. Fig. \ref{alpha_gamma_power} illustrates the comparative power intensity of the Alpha and Gamma bands across channels.

{\subsubsection{Eye Tracking Data}
An automated pipeline was developed to extract Alpha and Gamma-band EEG power synchronized with Eye-movement events from the SEED-IV dataset. First, all EEG `.mat` files were parsed to identify the correct 62-channel matrix, independent of specific MATLAB struct formatting. The signals were subsequently converted into MNE-Python `Raw` objects and filtered within the 1–50 Hz range. Simultaneously, Eye-tracking `.mat` files were processed using a generalized loader designed to handle nested arrays and inconsistent field names. All event onset times were translated into sample indices and verified to ensure sufficient pre- and post-stimulus data availability.

To align neural activity with visual behavior, EEG data were epoch-aligned to time windows of -0.5 s to +0.8 s relative to event onsets. Prior to spectral analysis, baseline correction was applied to remove non-task-related drifts. For a signal $x(t)$, the baseline-corrected signal $x_{corr}(t)$ is defined as \cite{handy2005event}:
\begin{equation}
x_{corr}(t) = x(t) - \frac{1}{T_{base}} \int_{t_{start}}^{t_{end}} x(\tau) \, d\tau
\label{eq:baseline_correction}
\end{equation}
where $[t_{start}, t_{end}]$ represents the baseline interval (e.g., -0.5 s to 0 s).

Following preprocessing, band power was computed for Alpha (8–14 Hz) and Gamma (31–50 Hz) bands. To extract the instantaneous power envelope, we utilized the Hilbert transform \cite{le2001comparison, freeman2007hilbert}. For a band-pass filtered signal $x_f(t)$, the analytic signal $z(t)$ is constructed as:
\begin{equation}
z(t) = x_f(t) + j \mathcal{H}[x_f(t)]
\label{eq:analytic_signal}
\end{equation}
where $\mathcal{H}[\cdot]$ denotes the Hilbert transform. The instantaneous power $P_{inst}(t)$ is then derived from the squared magnitude of the analytic signal:
\begin{equation}
P_{inst}(t) = |z(t)|^2 = x_f(t)^2 + (\mathcal{H}[x_f(t)])^2
\label{eq:instantaneous_power}
\end{equation}
The resulting mean band power, $\bar{P}$, for each channel is obtained by averaging the instantaneous power over the epoch duration $T$ \cite{sanei2013eeg}:
\begin{equation}
\bar{P} = \frac{1}{T} \int_{0}^{T} P_{inst}(t) \, dt
\label{eq:eye_mean_power}
\end{equation}
Eye-tracking raw signals underwent identical epoching and band-power extraction procedures. The final output of the pipeline was a structured, 62-channel dataset of Eye power features. These features form the basis for analyzing the relationship between Alpha–Gamma activity and oculomotor behaviors such as fixation, saccades, and pupil dilation. Notably, complete fixation, saccade, and pupil data were available for only 10 participants; therefore, this study restricts the eye-tracking analysis to this subset. This synchronization of EEG rhythms with precise visual events facilitates the direct linkage of cognitive states to visual attention behaviors, enabling more robust identification of deep-focus neural signatures.

\subsection{Model Classification with Explainability}

This phase of the methodology focuses on determining whether EEG- and Eye-tracking–derived features contain sufficient discriminative information to separate Alpha and Gamma activity patterns. This phase consists of two components: a classification step to evaluate model performance across multiple learning algorithms, and an explainability step using LIME to understand which channels or features most strongly drive the model’s decisions.

\subsubsection{Classification}
To rigorously evaluate the discriminative capacity of the extracted features, we implemented a classification framework designed to distinguish between Alpha- and Gamma-dominant cognitive states. This analysis utilized mean power intensity metrics derived from both EEG signals and Eye-tracking components (Fixation, Saccade, and Pupil). To ensure a comprehensive assessment, we benchmarked performance across a broad spectrum of architectures. Validated models included traditional machine learning algorithms, such as Logistic Regression \cite{hosmer2013applied}, Decision Tree \cite{de2013decision}, Random Forest \cite{rigatti2017random}, AdaBoost \cite{schapire2013explaining}, XGBoost \cite{ramraj2016experimenting}, and Gradient Boosting \cite{natekin2013gradient} as well as specialized deep learning models like CNN \cite{o2015introduction}, EEGNet \cite{lawhern2018eegnet}, and EEGFormer \cite{wan2023eegformer}. In this procedure, features from each modality were independently input into the classifiers to predict the cognitive state (Alpha vs. Gamma). This approach facilitated a rigorous comparative analysis of feature robustness across the different physiological signals.



\subsubsection{LIME Explainability}


 Local Interpretable Model-Agnostic Explanations (LIME) \cite{ribeiro2016should} is utilized to analyze the impact of EEG and Eye-tracking features (fixation, saccade, pupil) on Alpha and Gamma state classification. LIME produces local surrogate models that simulate the classifier's behavior in proximity to a particular input sample. This enables us to ascertain which EEG channels most significantly impact the model's predictions for each event type and frequency band.

\textbf{\textit{Local Perturbation and Neighborhood Construction:}} For a given input sample x, LIME produces a collection of perturbed samples \cite{ribeiro2016should}:
\begin{equation}
    \{x_1', x_2', \ldots, x_N'\}\label{eq}
\end{equation}




For EEG data, perturbations are applied to channel-wise signal features to analyze variations in the classifier’s output. Similarly, for eye-tracking data, perturbations are applied to channel-wise eye-movement features-fixation, saccadic, and pupil-related measures to analyze their effect on the classifier’s output.
To enforce locality, a proximity score is computed for each perturbed instance and used to weight the samples based on their distance from the original input in the feature space, assigning higher importance to perturbations that remain closer to the original sample \cite{ribeiro2016should}.

\begin{equation}
\pi(x, x_i') = \exp\left( -\frac{D(x, x_i')^2}{\sigma^2} \right)\label{eq}
\end{equation}

where, $\pi(x, x_i')$ refers to how close a perturbed sample $x_i'$ is to the original instance $x$. \( D(\cdot) \) represents the distance metric (Euclidean). \( \sigma \) regulates the locality of the neighborhood.

\textbf{\textit{Local Surrogate Model:}} LIME constructs a straightforward, interpretable model g(x) (usually linear regression) inside the altered vicinity to approximate the intricate classifier f(x) \cite{ribeiro2016should}.
\begin{equation}
g = \arg\min_{g \in G} \left( L(f, g, \pi) + \Omega(g) \right)\label{eq}
\end{equation}


where, $f$  denotes the original black-box classifier whose prediction is to be explained. $g$ represents the local surrogate model, typically a simple linear model, trained to approximate $f$ in the neighborhood of $x$.
 $\mathcal{G}$ represents the class of interpretable models. 
 $\pi$ denotes a proximity kernel that assigns higher weights to perturbed samples closer to $x$, thereby enforcing locality.
 $L(f,g,\pi)$ represents a local fidelity loss function that measures how well the surrogate model $g$ approximates the predictions of the black-box model $f$ in the vicinity of $x$.    \( \Omega(g) \) denotes interpretability (e.g., sparsity). \( g \) represents the category of interpretable models \cite{ribeiro2016should}.
This generates feature weight as mentioned below.
\begin{equation}
w_j = \frac{\partial g(x)}{\partial x_j}\label{eq}
\end{equation}
where, $g(x)$ is the surrogate model. $x_j$ denotes the j-th feature of the input instance. $w_j$ represents the weight of the feature importance of the j-th feature. Therefore, we can get $w_j$ from the derivative of $g(x)$ and $x_j$.

\textbf{\textit{Feature Importance Computation:}} The local linear model's coefficients, $w_j$, indicate the importance of each EEG feature or channel in classifying instances, particularly in relation to Alpha and Gamma Power intensity. For each sample, feature contributions are derived from eye-movement measures, including Alpha and Gamma fixation, saccade, and pupil dynamics. The analysis identifies the top 30 most significant channels by averaging channel-specific feature weights, with bar plots illustrating the brain regions that reflect changes in Eye movements as cognitive load varies.

\textbf{\textit{Interpretation:}} LIME analyzes EEG channels by ranking them based on their average absolute importance, highlighting the 30 channels that significantly influence Alpha and Gamma bands. The Alpha-band map corresponds to relaxed states, while the Gamma-band map reveals active frontal and temporal electrodes during intense focus. This analysis clarifies how brain regions impact classifier decisions, providing insights into eye movement variations and cognitive attention mechanisms, with spatial patterns reflecting neurophysiological findings.

Up to this point, classification and LIME explainability have been used to examine whether machine-learning models can reliably distinguish Alpha from Gamma activity and to determine which EEG channels or Eye-tracking features most influence these classifications. While these findings offer insight into feature importance from a predictive standpoint, they do not yet reveal the true spatial or oscillatory characteristics of the brain during deep cognitive focus. To address this core objective, the next phase of the methodology performs a detailed power intensity analysis and constructs topographical maps to localize the cortical regions exhibiting the strongest activity during Gamma-dominant states. These analyses play a crucial role in uncovering the neural patterns underlying deep cognition, moving beyond classifier behavior to physiologically grounded interpretation.

\subsection{Power Intensity and Topographical Map}

Power intensity combined with topographical analysis is used to characterize the cortical regions exhibiting heightened activity under Alpha- and Gamma-dominant conditions. Because higher band-specific power in a channel reflects stronger oscillatory engagement of the underlying brain region, analyzing channel-wise power provides a direct indication of spatial involvement in different cognitive states. The SEED-IV dataset provides EEG and Eye-tracking recordings from 45 participants, where Alpha and Gamma bands are extracted from 62 channels using a Butterworth bandpass filter with forward–backward processing to avoid phase distortion. Band power is then computed by averaging the squared EEG voltage over time, since instantaneous power is proportional 
$v(t)^2$, yielding the final power measure in µ$V^2$. The detailed procedure of power extraction and calculation is provided in Section II-Subsection A.

A topographical map \cite{duffy1989topographic} provides a two-dimensional representation of how EEG activity is spread out over the head.   Putting power levels for each channel on a head-shaped arrangement shows which parts of the brain have more or less spectral activity.   This spatial representation helps us understand how different mental states are connected to brain patterns.
The details of the establishment of the topographical map model for EEG data and Eye tracking data are mentioned below.

\textbf{\textit{Loading the feature and cleaning channel labels:}}
EEG and Eye tracking data features were standardized by cleaning channel names to align with the 10–20 \cite{mecarelli2019electrode} electrode naming convention. This normalization ensured consistency across the 62-channel montage, facilitating accurate spatial mapping and visualization in subsequent analysis steps.

\textbf{\textit{Selecting the standardized 62-channel montage:}}
In this step, all the 62 channels' names are provided to the model. 
These channel names correspond to the 10–20 montage system, specifying which electrodes are used and where they are positioned on the scalp.


\textbf{\textit{Computing mean Alpha and Gamma power per channel:}}
For the EEG and Eye tracking dataset, the Alpha and Gamma power per channel are imported from the dataset. After that, the mean of Alpha and Gamma power is calculated.

\textbf{\textit{Assigning 2-D scalp coordinates to electrodes:}} In this stage, each electrode in the 62-channel montage was assigned a 2-D coordinate (x, y) indicating its position on a schematic head. Frontal electrodes (FP1, FPZ, FP2) are at the top of the map, temporal electrodes (T7, T8) are on the left and right, respectively, and parietal or occipital electrodes (P, PO, O1, OZ, O2) are on the bottom. This coordinate dictionary provides the spatial information that is necessary to convert discrete electrode values into a continuous scalp topography.

\textbf{\textit{Interpolating channel values into a continuous topography:}} The model visualizes mean power values for each channel under specific conditions (Alpha or Gamma) by employing a coordinate map to extract x-coordinates, y-coordinates, and power values. It creates a high-resolution grid (300x300) over the scalp, spanning from -1.2 to 1.2 in both dimensions. Utilizing cubic interpolation, it estimates power values at each grid point, thereby converting sparse electrode data into a continuous 2-D field that depicts band power distribution across the scalp.

\textbf{\textit{Rendering the EEG topographical map:}} The matplotlib ``imshow" function visualizes an interpolated grid using a ``turbo" color scale to represent power intensities. The plot features a circular head outline, a triangular nose for orientation, ear shapes for left/right distinction, and electrode markers as white circles outlined in black, labeled by channel names. To emphasize head and activity patterns, axis ticks and frames are removed. A color bar labeled Power (µV²) provides the absolute scale.

\textbf{\textit{Generating separate Alpha and Gamma topographical maps:}} The function generates EEG topographies for Alpha and Gamma Bands, saving them as high-resolution PNGs. For eye-tracking data, two topographical maps are produced, showing fixation, Saccade, and Pupil Data, enabling visual comparison of neural activity patterns across different cognitive states (Alpha and Gamma).

\section{Experimental Results and Discussion}

\subsection{Dataset Description}

The SEED-IV dataset \cite{zheng2018emotionmeter} comprises recordings from 15 healthy, right-handed participants (8 females) aged 20 to 24 years. To evaluate cross-session stability, each participant completed three sessions on separate days, resulting in a total of 45 experimental recordings. The experimental protocol involved viewing 24 emotion-eliciting video clips per session, balanced across four emotional categories: happy, sad, fear, and neutral. Each clip, approximately two minutes in duration, was preceded by a 5-second cue. Successful emotion elicitation was validated through participant self-assessment using the Positive and Negative Affect Schedule (PANAS) \cite{thompson2007development} immediately following each viewing. The stimuli were selected from a larger pool of 168 clips rated by 44 independent evaluators; a final set of 72 clips with consistent arousal–valence ratings was distributed across the three sessions without repetition.

EEG signals were acquired using a 62-channel system with an international 10-20 electrode montage sampled at 1000 Hz. This specific montage was selected based on prior evidence suggesting that temporal regions are critical for emotional processing and suitable for wearable integrations. Simultaneously, eye-movement data was collected using Semantic Gaze Mapping (SMI) Eye Tracking Glasses (ETG). The system recorded raw parameters including pupil diameter, fixation duration and dispersion, saccade amplitude and duration, blink metrics, and event statistics.

The provided dataset includes standard signal preprocessing to ensure data quality. Raw EEG signals were band-pass filtered (1–75 Hz) to eliminate slow drifts and high-frequency noise, then resampled to align with the temporal resolution of the eye-tracking data. Artifacts unrelated to emotional processing such as environmental noise and minor electrode shifts were attenuated using a linear dynamic system (LDS) filter. Furthermore, the dataset categorizes neural activity into five distinct frequency bands: Delta (1–4 Hz), Theta (4–8 Hz), Alpha (8–14 Hz), Beta (14–31 Hz), and Gamma (31–50 Hz). For eye-tracking data, a Principal Component Analysis (PCA)-based method was applied to decouple ambient luminance effects from pupil diameter. By subtracting the first principal component, light-reflex artifacts were removed, isolating pupil responses specifically related to emotional modulation. Trials identified as emotionally inconsistent or mislabeled based on PANAS responses were excluded from analysis.

\subsection{Classification}
The classification task has been performed separately on EEG and Eye-tracking features to discriminate between Alpha and Gamma activities. Both modalities were evaluated using multiple machine-learning and deep-learning models via a 5-fold cross-validation protocol.


    \begin{table}[htbp]
\caption{Alpha vs Gamma EEG Classification}
\vspace{-0.4cm}
\begin{center}
\begin{tabular}{|
>{\centering\arraybackslash}p{1.3cm}|
>{\centering\arraybackslash}p{1.1cm}|
>{\centering\arraybackslash}p{1.1cm}|
>{\centering\arraybackslash}p{0.8cm}|
>{\centering\arraybackslash}p{0.7cm}|
>{\centering\arraybackslash}p{1cm}|
}

\hline
\cellcolor{cyan!10}\textbf{Classifier}&\cellcolor{cyan!10}\textbf{Accuracy}&\cellcolor{cyan!10}\textbf{Precision}&\cellcolor{cyan!10}\textbf{Recall}&\cellcolor{cyan!10}\textbf{F1-score}&\cellcolor{cyan!10}\textbf{Inference Time (ms)} \\
\hline
Gradient Boosting & \cellcolor{green!10}\textbf{\centering 90.0\%}& \cellcolor{green!10}\textbf{ \centering 91.92\%}& 88.8\%& \cellcolor{green!10}{\textbf{\centering 89.8\%}}&0.00089 \\
\hline
AdaBoost & 88.8\% & 87.92\%& 91.1\%& 89.2\%& 0.355\\
\hline

Logostic Regression & 85.5\% & 84.09\%& \cellcolor{green!10}\textbf{\centering 91.11\%}& 86.64\%&0.0032 \\
\hline

XGBoost & 85.56\%& 85.5\% &86.67\%& 85.82\% & 0.0312 \\
\hline
RF & 85.56\%& 85.47\%& 86.6\%& 85.6\%&0.1372 \\
\hline
Decision Tree &82.2\% & 82.05\% &84.4\%  &82.8\% & 0.0033 \\
\hline
CNN & 77.7\%& 77.9\%& 80\%& 78.07\%&2.53\\
\hline

EEGNet &  44.4\%& 36\% &17.7\%& 22.47\% & 4.1771\\
\hline
EEGFormer & 72.22\% & 73.34\%& 80.0\% & 74.7\% & 4.485\\
\hline
\end{tabular}
\label{EEG_class}
\end{center}
\end{table}

\subsubsection{\textit{EEG Classification}} After classifying Alpha vs Gamma using EEG data, it is found from Table \ref{EEG_class} that the Gradient Boosting is the best performing model with Accuracy of 90\%, Precision of 91.92\%, F1-score of 89.8\%, and inference time of 0.00089 ms.
In this work, the Gradient Boosting model was configured with a logistic loss, 100 decision trees, and a learning rate of 0.1, balancing learning stability and model complexity. Each tree had a maximum depth of 3 and was trained using the Friedman MSE splitting criterion, with all samples used at each boosting stage.
The best value of Recall is 91.11\%, and it is achieved in 0.0023 ms inference time using the Logistic Regression algorithm. In this study, the Logistic Regression model was trained in primal form with a convergence tolerance of 1e-4 and a maximum of 100 iterations. An intercept term was included, with no class reweighting and no elastic-net mixing. Training was performed without warm starts, verbose output was disabled, and no parallelization was specified. 


\begin{table}[t]
\centering
\scriptsize   
\caption{Alpha vs Gamma Eye Tracking Classification}
\vspace{-0.1cm}
\setlength{\tabcolsep}{3pt}
\begin{tabular}{|c|c|c|c|>{\centering\arraybackslash}p{0.8cm}|>{\centering\arraybackslash}p{1cm}|}
\hline
\cellcolor{cyan!10}\textbf{Model} &\cellcolor{cyan!10}\textbf{Accuracy} & \cellcolor{cyan!10}\textbf{Precision} & \cellcolor{cyan!10}\textbf{Recall} & \cellcolor{cyan!10}\textbf{F1-Score} & \cellcolor{cyan!10}\textbf{Inference Time (ms)} \\
\hline

\multirow{3}{*}{Logistic Regression}
& 70.0\% & 68.3\% & 70.0\% & 66.0\% & 0.0151 \\
& 80.0\% & 86.7\% & 80.0\% & 78.7\% & 0.0144 \\
& 80.0\% & 86.7\% & 80.0\% & 78.7\% & 0.0154 \\
\hline

\multirow{3}{*}{Decuision Tree}
& 80.0\% & 86.7\% & 80.0\% & 78.7\% & 0.0124 \\
& 75.0\% & 80.0\% & 75.0\% & 74.0\% & 0.0119 \\
&\cellcolor{green!10}\textbf{85.0\%} & \cellcolor{green!10}\textbf{90.0\%}& \cellcolor{green!10}\textbf{85.0\%} & \cellcolor{green!10}\textbf{84.0\%} & 0.0128 \\
\hline

\multirow{3}{*}{Random Forest}
& \cellcolor{green!10}\textbf{90.0\%} &\cellcolor{green!10} \textbf{93.3\%} & \cellcolor{green!10}\textbf{90.0\%} & \cellcolor{green!10}\textbf{89.3\%} & 0.6223 \\
& \cellcolor{green!10}\textbf{90.0\%} & \cellcolor{green!10}\textbf{93.3\%} & \cellcolor{green!10}\textbf{90.0\%} & \cellcolor{green!10}\textbf{89.3\%} & 0.6093 \\
& 85.0\% & 90.0\% & 85.0\% & 84.0\% & 0.6505 \\
\hline

\multirow{3}{*}{AdaBoost}
& 80.0\% & 78.3\% & 80.0\% & 76.0\% & 0.9439 \\
& 85.0\% & 90.0\% & 85.0\% & 84.0\% & 0.6296 \\
& 85.0\% & 90.0\% & 85.0\% & 84.0\% & 0.7594 \\
\hline

\multirow{3}{*}{Gradient Boosting}
& 80.0\% & 86.7\% & 80.0\% & 78.7\% & 0.0310 \\
& 85.0\% & 90.0\% & 85.0\% & 84.0\% & 0.0402 \\
& 85.0\% & 90.0\% & 85.0\% & 84.0\% & 0.0402 \\
\hline

\multirow{3}{*}{XGBoosting}
& 70.6\% & 68.9\% & 75.0\% & 71.8\% & 0.1400 \\
& 85.0\% & 90.0\% & 85.0\% & 84.0\% & 0.1394 \\
& 85.0\% & 90.0\% & 85.0\% & 84.0\% & 0.1545 \\
\hline

\multirow{3}{*}{Convolutional
Neural
Network}
& 60.0\% & 58.3\% & 60.0\% & 56.0\% & 11.5520 \\
& 65.0\% & 61.7\% & 65.0\% & 60.0\% & 11.2554 \\
& 75.0\% & 76.7\% & 75.0\% & 74.7\% & 14.3922 \\
\hline

\multirow{3}{*}{EEGNet}
& 60.0\% & 48.3\% & 60.0\% & 49.3\% & 19.3025 \\
& 55.0\% & 46.7\% & 55.0\% & 48.0\% & 19.3116 \\
& 50.0\% & 40.0\% & 50.0\% & 42.0\% & 25.3187 \\
\hline

\multirow{3}{*}{EEGFormer}
& 65.0\% & 65.0\% & 65.0\% & 60.7\% & 19.9612 \\
& 70.0\% & 73.3\% & 70.0\% & 69.3\% & 20.6422 \\
& 70.0\% & 73.3\% & 70.0\% & 69.3\% & 21.5216 \\
\hline

\end{tabular}
\label{tab:eye_classification}
\end{table}

\subsubsection{\textit{Eye Tracking Classification}} After classifying Alpha vs Gamma using Eye
 tracking data, it is found from Table \ref{tab:eye_classification} that the Random Forest is the best performing model for Fixation and Saccade. The Accuracy, Precision, Recall, and F1-score are 90\%, 93\%, 90\%, 89.33\%, respectively. 
In this analysis, the Random Forest model was implemented with 100 decision trees using the Gini impurity criterion. Each tree had no depth limit and followed standard splitting rules, such as a minimum of 2 samples per split and a minimum of 1 sample per leaf. At each split, a subset of features equal to the square root of the total features was considered. Bootstrap sampling was enabled to generate diverse trees, while no out-of-bag scoring, class weighting, or pruning was applied.

 For pupil, the Decision tree is the best performing model, and its Accuracy, Precision, Recall, and F1-score are 85\%, 90\%, 85\%, 84\%, respectively. 
In this study, the Decision Tree model was implemented using the Gini impurity criterion with the best-split strategy. The tree depth was left unrestricted to allow full growth, while node splitting required at least two samples, and leaf nodes contained at least one sample.

Therefore, using a Random Forest, the best accuracy achieved was 90\% for fixation and 90\% for Saccade, with inference times of 0.6223 and 0.6093 ms, respectively. 
Furthermore, using the Decision Tree, 85\% accuracy is achieved for pupils with an inference time of 0.0128 ms. In the table \ref{tab:eye_classification} for each model, three values are
reported: fixation in the first, saccade in the second, and
pupil in the third row.  


\subsection{LIME Explainability}
After classification, the best model for EEG (Gradient Boosting) and Eye tracking data (Random Forest: Fixation and Saccade; Decision Tree: Pupil) are used for LIME Explainer.
LIME explainability is used to identify which EEG channels contribute most to the model’s predictions by measuring their local impact on classification outcomes. This helps interpret the decision-making process of the model and highlights the specific brain regions influencing each frequency band. The following subsections describe the influential channels for EEG and Eye-tracking data separately.

\subsubsection{\textit{EEG Analysis}} As shown in Fig. \ref{alpha_gamma_LIME}, the orange colored channels or electrodes are positively influencing the Gradient Boosting model to predict the Gamma band. The list of positively influencing channels are C4 (Central Right), POZ (Parieto-Occipital Midline), T8 (Temporal Right), F5 (Frontal Left), C1 (Central Left), F4 (Frontal Right), CB1 (Mastoid Left), PO8 (Parieto-Occipital Right), P6 (Parietal Right), TP7 (Temporo-Parietal Left), FT8 (Fronto-Temporal Right), Fp1 (Frontopolar Left), CP4 (Centro-Parietal Right), PO3 (Parieto-Occipital Left), CP1 (Centro-Parietal Left), and OZ (Occipital Midline).   The blue colored electrodes are influencing the model to predict the Alpha band.
\begin{figure}[htbp]
\centerline{\includegraphics[width=0.5\textwidth, trim=0 0 0 0,clip]{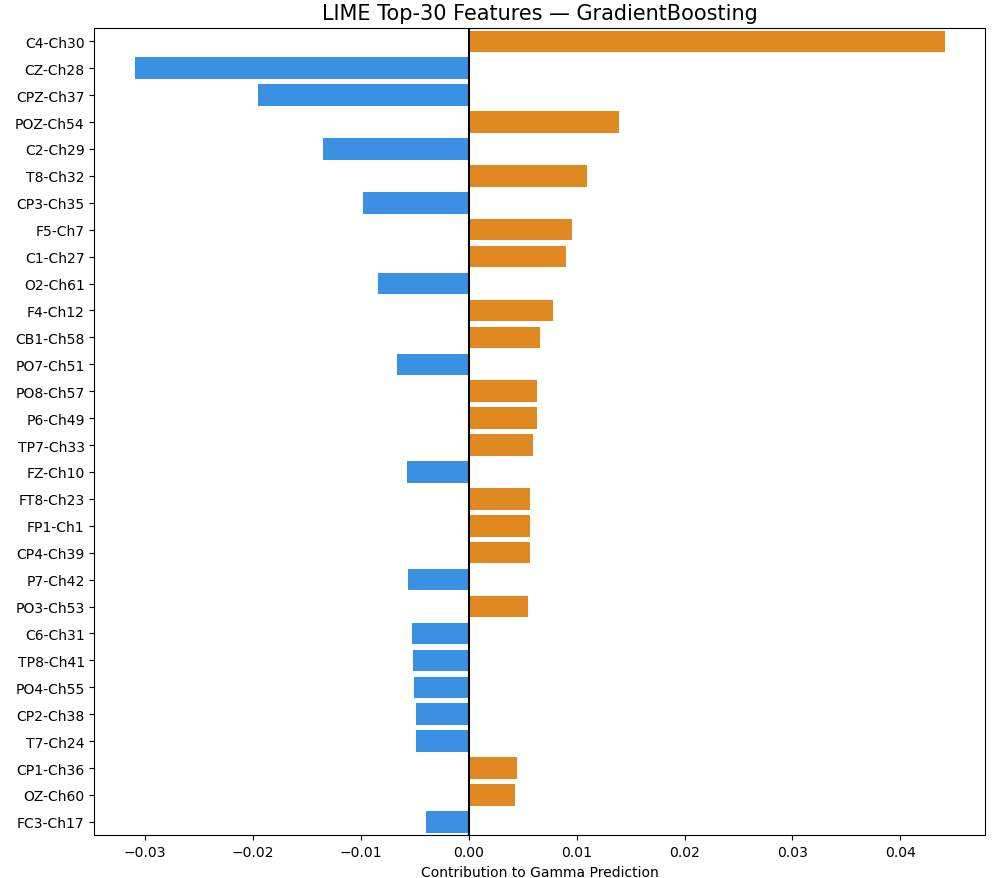}}
\vspace{-0.3cm}
\caption{LIME explainability using EEG data.}
\label{alpha_gamma_LIME}
\end{figure}

\begin{figure*}[htbp]
\centerline{\includegraphics[width=1\textwidth, trim=0 0 0 58,clip]{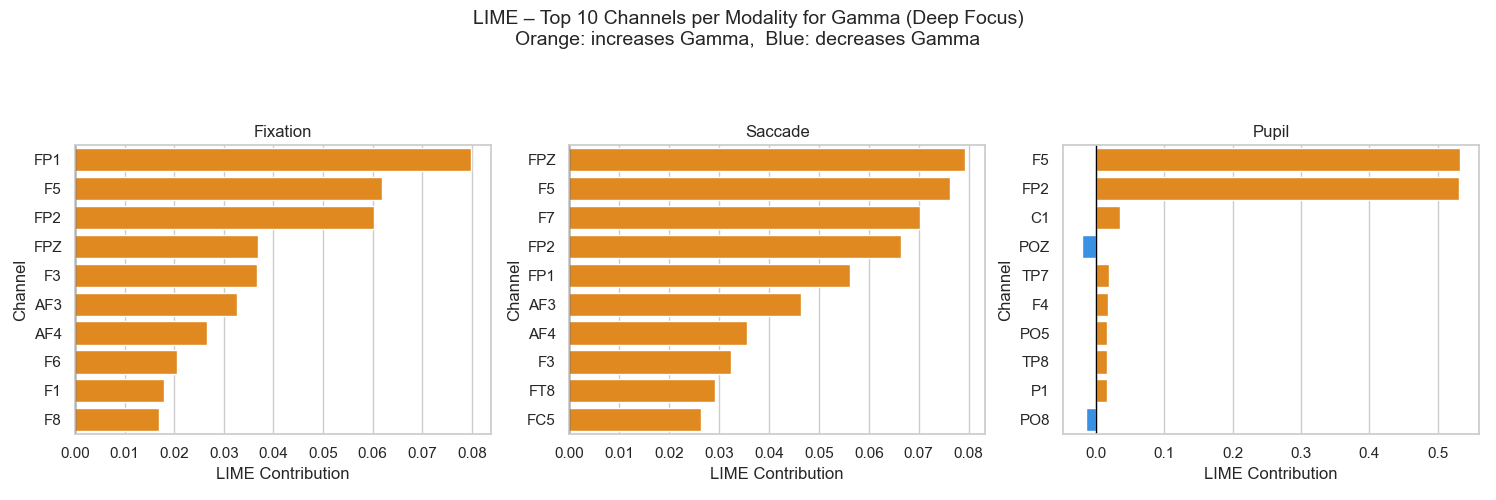}}
\vspace{-0.5cm}
\caption{LIME explainability using Eye tracking data.}
\label{eye_gamma_lime}
\end{figure*}
\subsubsection{\textit{Eye Tracking Analysis}} Using LIME, we can identify the features that are important for Gamma vs Alpha classification. According to the Fig. \ref{eye_gamma_lime}, for Fixation, channel or electrode FP1 (Frontopolar Left), FP2 (Frontopolar Right), F5 (Frontal Left), F3 (Frontal Left), FPZ (Frontopolar Midline), AF3 (Anterior Frontal Left), F6 (Frontal Right), F1 (Frontal Midline-Left), F8 (Frontal Right), and AF4 (Anterior Frontal Right) positively influence the classification of the Gamma band. For Saccade, channel FPZ (Frontopolar Midline), F5 (Frontal Left), AF3 (Anterior Frontal Left), F3 (Frontal Left), FP1 (Frontopolar Left), F7 (Frontal Left Lateral), FT8 (Fronto-Temporal Right), AF4 (Anterior Frontal Right), FP2 (Frontopolar Right), and FC5 (Fronto-Central Left) exhibit a positive contribution to Gamma-band prediction. Furthermore, for Pupil electrode F5 (Frontal Left), FP2 (Frontopolar Right), C1 (Central Left), TP7 (Temporo-Parietal Left), F4 (Frontal Right), PO5 (Parieto-Occipital Left), TP8 (Temporo-Parietal Right), and P1 (Parietal Midline-Left) are influencing positively and POZ and PO8 is influencing negatively for Gamma band (influencing positively for Alpha band). 

Classification and LIME explainability are used to assess the ability of machine-learning models to differentiate Alpha- and Gamma-dominant activity and to identify influential EEG channels or Eye-tracking features. However, understanding the oscillatory characteristics associated with deep cognitive focus requires direct analysis of power intensity. Accordingly, the next phase focuses on uncovering neural patterns related to deep cognition beyond classifier behavior.
\subsection{Power Intensity Analysis}

Power‑intensity analysis facilitates a more intuitive interpretation of signal strength across EEG electrodes or channels.

\subsubsection{\textit{Power Analysis of EEG Data}}
To assess the oscillatory strength of neural activity across the cortex, we calculated the mean power for EEG channels within Alpha and Gamma bands. Each of the 62-channel raw EEG recordings underwent a 4th-order Butterworth band-pass filter to focus on the target frequency range. We determined power from the mean-square amplitude of the filtered signal, indicative of average signal energy per electrode. The mean values for Alpha and Gamma were averaged across corresponding trials, yielding stable estimates of Mean Power for each channel, thus mapping the spatial oscillatory strength of cortical regions.

Moreover, we analyzed burst-based features to capture the transient characteristics of neural oscillations. A burst is defined as a short period where oscillatory amplitude significantly surpasses baseline levels, indicating increased neural synchrony. After filtering, we computed the instantaneous amplitude envelope using the analytic Hilbert transform. Bursts were identified when the envelope exceeded a statistical threshold set at \( (\mu + 2\sigma) \), with \( \mu \)  and \( \sigma \) being the mean and standard deviation of the envelope. We extracted three key temporal features from these bursts: (1) Burst Count, the total number of unique burst episodes; (2) Burst Duration, the average time above threshold in milliseconds; and (3) Burst Rate, the frequency of bursts per second. These metrics provide additional insight into the frequency and duration of high-amplitude neural oscillations in the Alpha and Gamma bands.
Table \ref{tab:alpha_gamma_meanpower_burst_top10} shows the top 10 channels and their mean power, burst rate, burst count, and burst duration of the Alpha and Gamma bands.
As shown in the table, the Gamma band exhibits maximum power at channel T8 (46.69 µV²), whereas the Alpha band shows its highest power at channel FPZ (147.37 µV²). Another noticeable observation is that the Burst rate, count, and duration of Alpha bands are substantially lower than those observed in the Gamma band.
Apart from the table, among all 62 channels, the CPZ channel has the minimum power of 0.089 µV² and 0.12 µV² for the Gamma and Alpha bands, respectively. After performing statistical analysis with all 62 channels' Mean power, it is found that the average of the Mean power is 5.99, and the Standard deviation is 7.00 for the Gamma band.
Similarly, for the Alpha band, the average of the Mean power is 16.47 µV², and the Standard deviation is 29.9.

\begin{table}[t]
\centering
\scriptsize
\caption{Top-10 Channels: Alpha and Gamma Mean Power and Burst Characteristics}
\vspace{-0.2cm}
\setlength{\tabcolsep}{2pt}
\renewcommand{\arraystretch}{1.9}
\begin{tabular}{
|>{\centering\arraybackslash}p{.9cm}|
 >{\centering\arraybackslash}p{.7cm}|
 >{\centering\arraybackslash}p{.7cm}|
 >{\centering\arraybackslash}p{.7cm}|
 >{\centering\arraybackslash}p{.7cm}|
 >{\centering\arraybackslash}p{.7cm}|
 >{\centering\arraybackslash}p{.7cm}|
 >{\centering\arraybackslash}p{.7cm}|
 >{\centering\arraybackslash}p{.7cm}|
}
\hline
\multirow{2}{*}{\textbf{Channel}} &
\multicolumn{4}{c|}{\cellcolor{cyan!10}\textbf{Alpha}} &
\multicolumn{4}{c|}{\textbf{\cellcolor{cyan!10}Gamma}} \\

\cline{2-9}
& \cellcolor{cyan!10}\textbf{Mean Power} & \cellcolor{cyan!10}\textbf{Burst Rate} & \cellcolor{cyan!10}\textbf{Burst Count} & \cellcolor{cyan!10}\textbf{Burst Duration} 
& \cellcolor{cyan!10}\textbf{Mean Power} & \cellcolor{cyan!10}\textbf{Burst Rate} & \cellcolor{cyan!10}\textbf{Burst Count} & \cellcolor{cyan!10}\textbf{Burst Duration}  \\
\hline
FPZ & 147.37 & 0.4577 & 17.18 & 124.10 & 22.69 & 1.2762 & 47.33 & 40.62 \\
\hline
FP1 & 140.25 & 0.4534 & 16.96 & 124.22 & 22.25 & 1.2623 & 46.89 & 40.34 \\
\hline
FP2 & 132.62 & 0.4629 & 17.36 & 117.76 & 20.79 & 1.2698 & 47.16 & 39.48 \\
\hline
AF4 & 55.81  & 0.4398 & 16.40 & 116.15 & 11.29 & 1.2510 & 46.53 & 36.61 \\
\hline
AF3 & 41.73  & 0.4403 & 16.42 & 118.97 & 9.87  & 1.2429 & 46.24 & 37.39 \\
\hline
T8  & 40.95  & 0.3739 & 13.93 & 110.17 & 46.69 & 1.1395 & 42.33 & 33.82 \\
\hline
F7  & 29.01  & 0.4021 & 14.98 & 115.24 & 8.56  & 1.2008 & 44.73 & 35.79 \\
\hline
F8  & 27.28  & 0.3917 & 14.58 & 112.24 & 7.53  & 1.2508 & 46.60 & 35.37 \\
\hline
PO5 & 22.71  & 0.3763 & 13.98 & 107.69 & 18.02 & 1.1536 & 42.67 & 35.29 \\
\hline
F6  & 17.52  & 0.3933 & 14.56 & 114.24 & 5.67  & 1.2183 & 45.31 & 35.71 \\
\hline
\end{tabular}
\label{tab:alpha_gamma_meanpower_burst_top10}
\end{table}

Based on the intensity or the strength of the signal per channel, Fig. \ref{alpha_gamma_feature} shows the most influential 30 channels for both Alpha and Gamma bands. From Fig. \ref{alpha_gamma_feature}, it is observed that the channel FPZ (Frontopolar Midline), FP1 (Frontopolar Left), FP2 (Frontopolar Right), AF4 (Anterior Frontal Right), AF3 (Anterior Frontal Left), T8 (Temporal Right), F7 (Frontal Left), F8 (Frontal Right), PO5 (Parieto-Occipital Left), and F6 (Frontal Right) are the top 10 most influential features. For clarity, both the channel names and their corresponding number are provided (e.g., FPZ-Ch2). 

\begin{figure}[htbp]
\centerline{\includegraphics[width=0.5\textwidth]{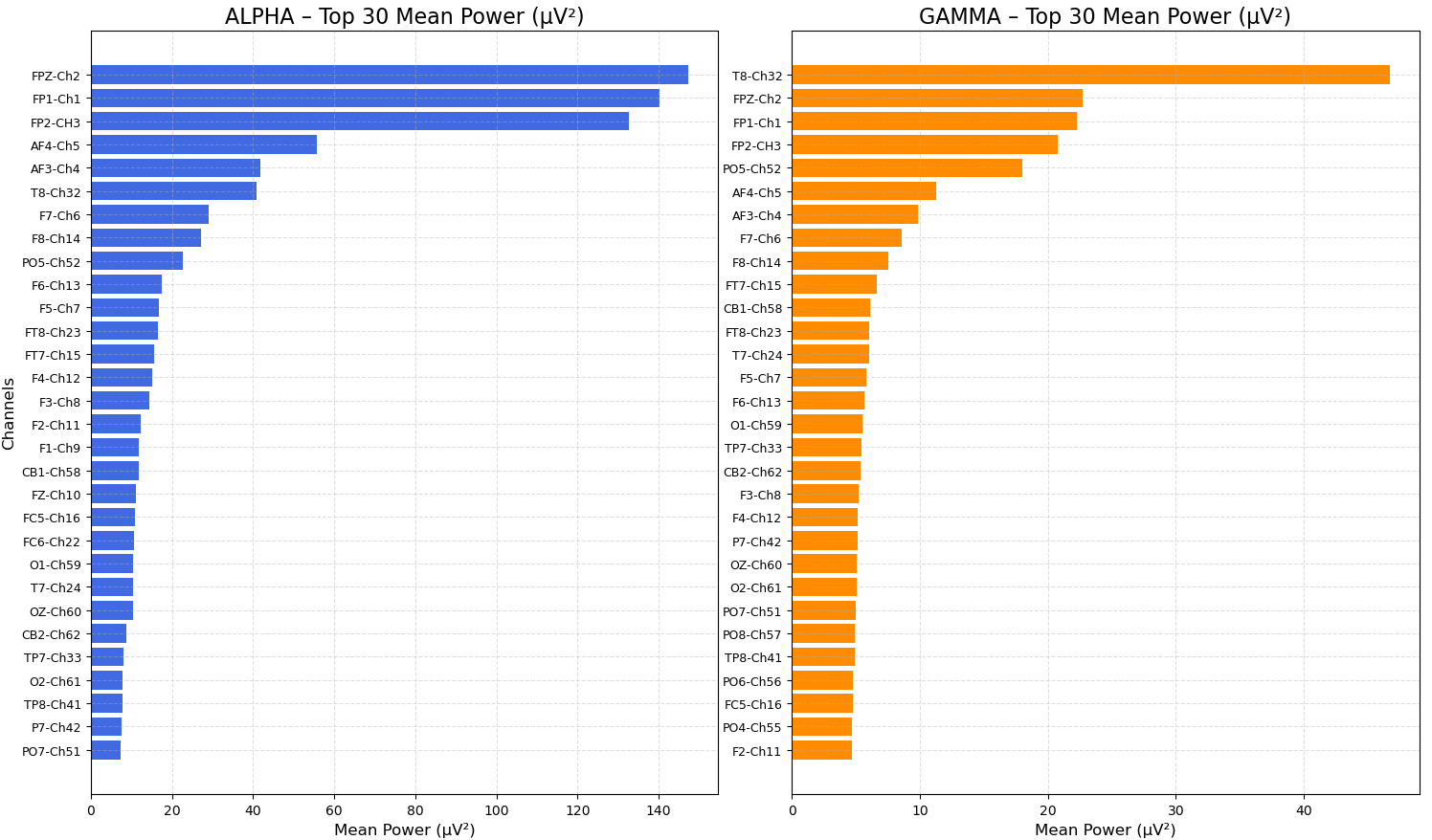}}
\vspace{-0.3cm}
\caption{Channels and their importance for Alpha and Gamma bands.}
\label{alpha_gamma_feature}
\end{figure}

For the Gamma band, the important channels and their level of intensity are also shown in Fig. \ref{alpha_gamma_feature}. Based on the strength of channel or the power, it can be said that T8 (Temporal Right), FPZ (Frontopolar Midline), FP1 (Frontopolar Left), FP2 (Frontopolar Right), PO5 (Parieto-Occipital Left), AF4 (Anterior Frontal Right), AF3 (Anterior Frontal Left), F7 (Frontal Left Lateral), F8 (Frontal Right Lateral), and FT7 (Frontotemporal Left) are the top 10 most important channnels. These channels or areas of the brain are producing the strongest Gamma activity. A channel with high Gamma power indicates that the brain region under that electrode is highly active and engaged in complex processing. So, it can be stated that these channels or these brain areas are highly responsible for the deep focus state.

\begin{table}[t]
\centering
\footnotesize
\caption{Average Fixation, Saccade, and Pupil Power for 10 Channels in Alpha and Gamma Bands}
\vspace{-0.2cm}
\setlength{\tabcolsep}{4pt}
\renewcommand{\arraystretch}{1.25}

\begin{tabular}{|
>{\centering\arraybackslash}p{.99cm}|
>{\centering\arraybackslash}p{.90cm}|
>{\centering\arraybackslash}p{.90cm}|
>{\centering\arraybackslash}p{.90cm}|
>{\centering\arraybackslash}p{.90cm}|
>{\centering\arraybackslash}p{.90cm}|
>{\centering\arraybackslash}p{.90cm}|
}

\hline
\multirow{2}{*}{{\textbf{Channel}} }
& \multicolumn{3}{c|}{\cellcolor{cyan!10}\textbf{Alpha}} 
& \multicolumn{3}{c|}{\cellcolor{cyan!10}\textbf{Gamma}} \\ 
\cline{2-7}
&  \cellcolor{cyan!10}\textbf{Fixation} 
& \cellcolor{cyan!10} \textbf{Saccade} 
& \cellcolor{cyan!10}\textbf{Pupil}
& \cellcolor{cyan!10} \textbf{Fixation} 
& \cellcolor{cyan!10}\textbf{Saccade} 
&  \cellcolor{cyan!10}\textbf{Pupil} \\
\hline

FP1  & 93.19 & 138.63 & 151.56 & 26.09 & 21.87 & 22.64 \\
\hline
FPZ  & 102.12 & 130.01 & 140.06 & 30.09 & 20.78 & 21.36 \\\hline
FP2  & 112.23 & 139.19 & 151.91 & 31.64 & 22.66 & 23.45 \\\hline
AF3  & 80.38 & 117.74 & 127.66 & 27.51 & 18.72 & 19.13 \\\hline
AF4  & 95.77 & 118.19 & 129.97 & 30.31 & 19.57 & 19.72 \\\hline
F7   & 66.05 & 104.46 & 114.24 & 22.78 & 16.95 & 17.41 \\\hline
F8   & 69.45 & 105.04 & 118.18 & 24.91 & 17.06 & 17.81 \\\hline
F6   & 55.21 & 90.79  & 98.15  & 20.11 & 14.93 & 15.43 \\\hline
FT7  & 62.30 & 100.30 & 112.71 & 22.00 & 16.30 & 16.74 \\\hline
CB1  & 8.99  & 14.63  & 15.78  & 4.28  & 2.88  & 3.12 \\
\hline

\end{tabular}
\label{tab:eye_alpha_gamma_power}
\end{table}

\begin{table*}[htbp]
\caption{Alpha vs Gamma Analysis Across EEG Channels}
\vspace{-0.7cm}
\begin{center}
\resizebox{\textwidth}{!}{%
\begin{tabular}{|
>{\centering\arraybackslash}p{1.5cm}|
>{\centering\arraybackslash}p{3.2cm}|
>{\centering\arraybackslash}p{4.2cm}|
>{\centering\arraybackslash}p{4.2cm}|
>{\centering\arraybackslash}p{4.2cm}|
}

\hline
\cellcolor{cyan!10}\textbf{Channel} &
\cellcolor{cyan!10}\textbf{Brain Region} &
\cellcolor{cyan!10}\textbf{Key Cognitive Functionality} &
\cellcolor{cyan!10}\textbf{Alpha Band Pattern} &
\cellcolor{cyan!10}\textbf{Gamma Band Pattern} \\
\hline

T8 & Right Temporal Cortex &
Sensory integration; auditory and emotional processing &
Low Alpha; temporal regions show weaker resting Alpha &
Very high Gamma; strongest activation for cognitive load and sensory integration \\
\hline

FPZ & Medial Prefrontal Cortex &
Executive control; planning; sustained attention &
Very high Alpha; relaxed attention; frontal disengagement &
High Gamma; strong involvement in executive control and decision making \\
\hline

FP1 & Left Prefrontal Cortex &
Working memory; emotional regulation; cognitive planning &
High Alpha; prefrontal relaxation during Label 0 &
Moderate Gamma; contributes to cognitive control \\
\hline

FP2 & Right Prefrontal Cortex &
Vigilance; attentional control; arousal regulation &
High Alpha; reduced frontal activation during Label 0 &
Moderate Gamma; active during attentional processes \\
\hline

PO5 & Visual--spatial Cortex &
Visual processing; spatial attention; mental imagery &
Moderate Alpha; strong occipital Alpha during rest &
High Gamma; strong activation during visual attention and cognitive engagement \\
\hline

AF4 & Right DLPFC &
Cognitive flexibility; task switching; attention control &
High Alpha; reduced cognitive load during resting &
Moderate to high Gamma; active during working memory and attentional tasks \\
\hline

AF3 & Left DLPFC &
Reasoning; analytical problem solving; working memory &
High Alpha; lower working memory demand &
Moderate Gamma; increases with analytical and reasoning effort \\
\hline

F7 & Inferior Frontal &
Language processing; inhibitory control; decision making &
Low to moderate Alpha &
Low Gamma; minimal activation in Gamma map \\
\hline

F8 & Inferior Frontal &
Emotional regulation; response inhibition; attention shifting &
Low to moderate Alpha &
Moderate Gamma; involved in emotional evaluation and inhibitory control \\
\hline

FT7 & Frontotemporal Junction &
Multisensory integration; language comprehension; verbal working memory &
Low Alpha; language regions typically show reduced Alpha &
Low to moderate Gamma; moderate involvement in multimodal processing \\
\hline

\end{tabular}
}
\end{center}
\label{comparison}
\end{table*}

\subsubsection{\textit{Power Analysis of Eye Tracking Data}} The mean values of fixation, saccade, and pupil power from an Eye-tracking dataset are calculated. Each record was labeled by frequency band (Alpha or Gamma). For the channels FP1, FPZ, FP2, AF3, AF4, F7, F8, F6, FT7, and CB1, data were divided by band label and averaged, resulting in six measures per channel: mean fixation, saccade, and pupil values for both Alpha and Gamma bands. This provides a stable estimate of Eye movement behavior related to Alpha and Gamma specific activity for each region of interest.
Table \ref{tab:eye_alpha_gamma_power} shows the mean power or the intensity of the signal for the top 10 channels.

Eye-tracking data, including eye-movement behavior during task performance, provide important information for analyzing brain activity. In this study, fixation, saccades, and pupil dilation are considered as eye tracking features. Fig. \ref{Eye_Alpha} presents the important channels based on the strength received from the electrodes (channels) for the Alpha band. It is observed that the FPZ (Frontopolar Midline), FP1 (Frontopolar Midline), FP2 (Frontopolar Right), AF3 (Anterior Frontal Left), AF4 (Anterior Frontal Left), F7 (Frontal Left), F8 (Frontal Right), F6 (Frontal Right), F5 (Frontal Left), and FT8 (Frontotemporal Right) are the top 10 most influential channels associated with eye fixation and saccade. For pupil-related measures, the same nine channels are identified, with the exception of the tenth channel, which is F4 (Frontal Right).

\begin{figure}[htbp]
\centerline{\includegraphics[width=0.5\textwidth, trim=60 0 0 60.5,clip]{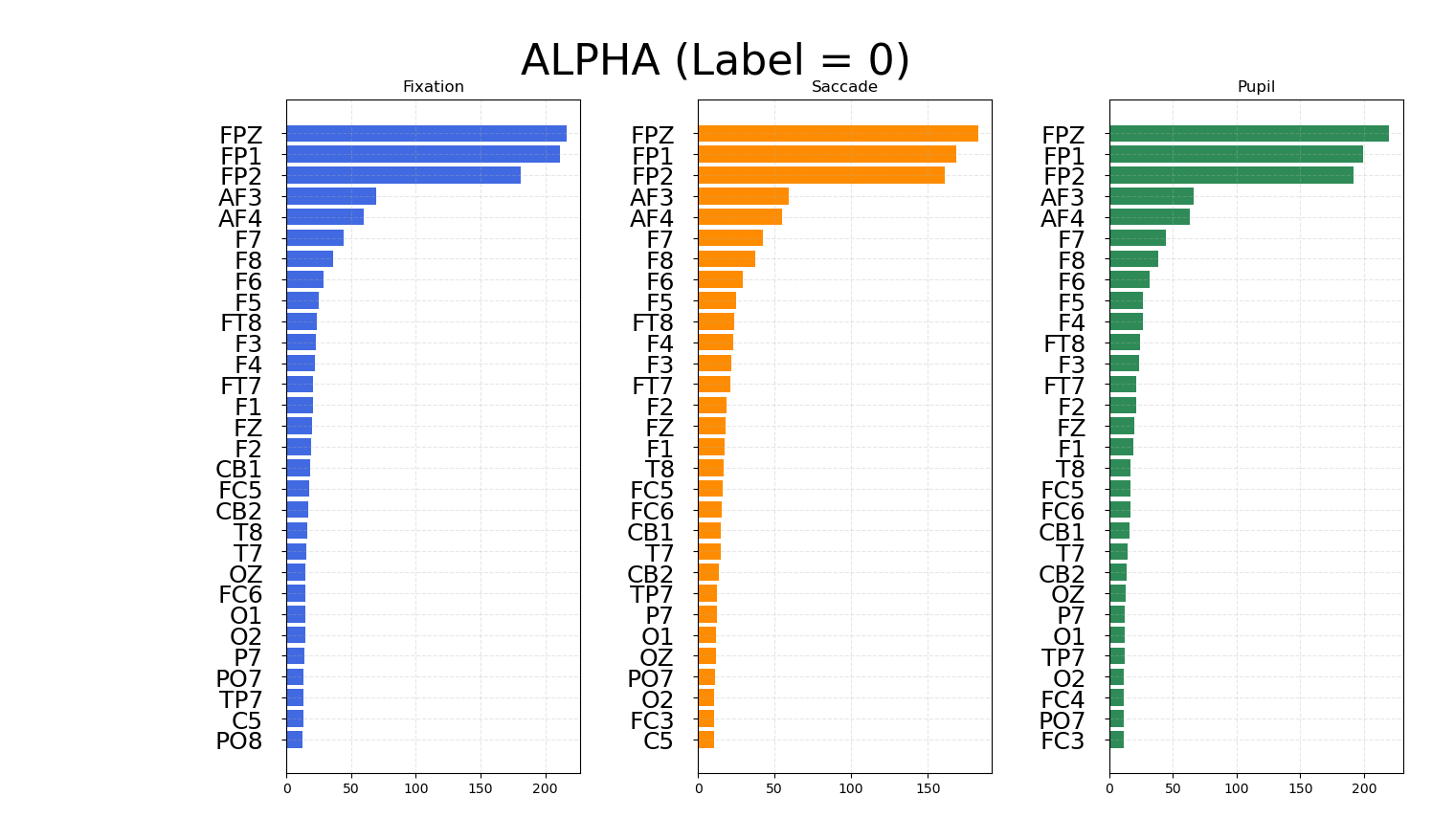}}
\vspace{-0.5cm}
\caption{Top 30 influential channels for eye tracking data during the Alpha Band. The X axis resembles the mean power, and the Y axis is the number of channels or electrodes. }
\label{Eye_Alpha}
\end{figure}

\begin{figure}[htbp]
\centerline{\includegraphics[width=0.5\textwidth, trim=60 0 0 61,clip]{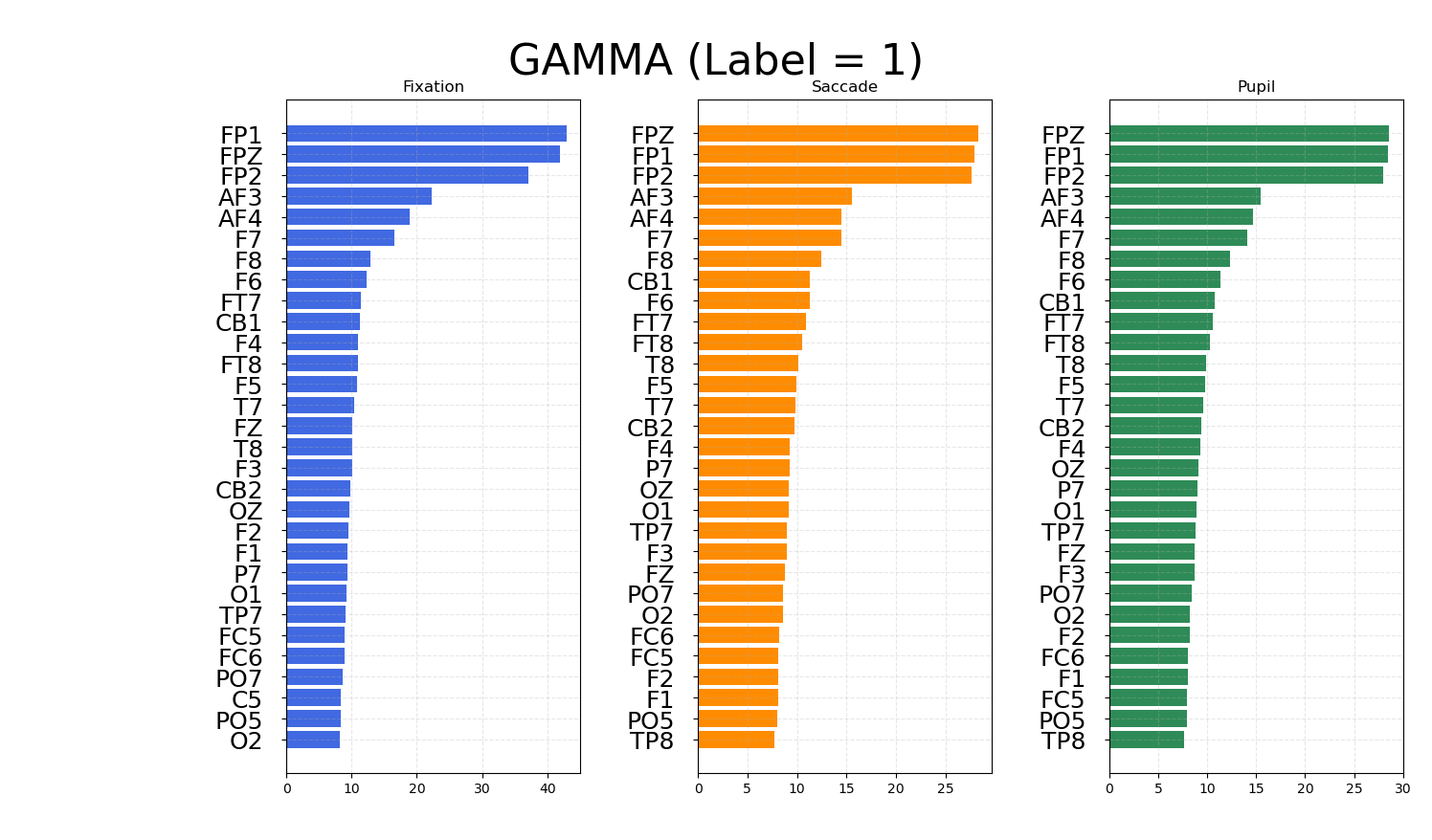}}
\vspace{-0.5cm}
\caption{Top 30 influential channels for eye tracking data during the Gamma band. The X axis resembles the mean power, and the Y axis is the number of channels or electrodes.}
\label{Eye_Gamma}
\end{figure}
Fig. \ref{Eye_Gamma} depicts the top 30 most influential channels or the region of the brain that has been extracted from the eye tracking data. Features, such as Fixation, Saccade, and Pupil Dilation, are used in this analysis. After analyzing the Gamma band's data it is found that the power strength of FP1 (Frontopolar Left), FPZ (Frontopolar Midline), FP2 (Frontopolar Right), AF3 (Anterior Frontal Left), AF4 (Anterior Frontal Right), F7 (Frontal Left), F8 (Frontal Right), F6 (Frontal Right), FT7 (Frontotemporal Left), and CB1 (Cerebellar Left) are the highest power strength. For Saccade and pupil, the top 10 channels that have the most power strength are also similar to the fixation, with only minor differences. Specifically, FP1 emerges as the second most influential channel for both saccade and pupil features, while CB1 ranks as the eighth most influential channel for saccade and pupil, compared to its tenth-rank position for fixation.

\begin{table*}[htbp]
\caption{Importance of EEG Channels for Eye-Tracking Related Cognitive Processing}
\vspace{-0.4cm}
\begin{center}
\begin{tabular}{|
>{\centering\arraybackslash}p{1.5cm}|
>{\centering\arraybackslash}p{3.5cm}|
>{\centering\arraybackslash}p{10.5cm}|
}

\hline
\cellcolor{cyan!10}\textbf{Channel} & \cellcolor{cyan!10}\textbf{Brain Region} &\cellcolor{cyan!10} \cellcolor{cyan!10}\textbf{Importance for Eye Tracking} \\
\hline

FP1 & Left Frontopolar Cortex &
Sensitive to blink artifacts and fixation stability; reflects attention switching and cognitive load during fixation tasks. \\
\hline

FPZ & Midline Frontopolar Cortex &
Involved in top-down attentional control and mental effort during gaze shifts, captures ocular artifacts (saccades). \\
\hline

FP2 & Right Frontopolar Cortex &
Responds to right-Eye movements and vigilance; useful for detecting attentional shifts and saccade-related activity. \\
\hline

AF3 & Left Anterior Frontal Cortex &
Reflects working memory and attention during fixation; captures early cognitive processing of visual attention. \\
\hline

AF4 & Right Anterior Frontal Cortex &
Involved in attention shifting and visual scanning behavior; linked to cognitive load during saccades. \\
\hline

F7 & Left Frontal Cortex &
Important for inhibiting unwanted Eye movements; involved in visual search behavior and gaze or focus-driven decision-making. \\
\hline

F8 & Right Frontal Cortex &
Tracks emotional gaze processing and inhibitory control; important for spatial gaze orientation. \\
\hline

F6 & Right Frontal Cortex &
Associated with high cognitive load during visual tracking; controls gaze and attentional focus during demanding tasks. \\
\hline

FT7 & Left Frontotemporal Junction &
Processes visual-spatial cues during gaze shifts; important for Eye movements in reading and language tasks. \\
\hline

CB1 & Left Cerebellum &
Critical for Eye-movement coordination; supports smooth pursuit, saccadic accuracy, and gaze stabilization during tracking. \\
\hline

\end{tabular}
\end{center}
\label{eye}
\end{table*}

\subsection{Topographical Map}
A topographical map is a 2D visualization that displays the spatial distribution of EEG activity across the scalp using color gradients, highlighting the relative activity of different brain regions or electrodes within a specific frequency band. The topographical maps for both EEG and eye-tracking data are discussed below.

\subsubsection{\textit{EEG Topographical Map}} Fig. \ref{Topo_Alpha_Gamma} (a) illustrates the EEG topographical map for the Alpha band. The map shows that the previously identified channels exhibit high power intensity, indicating that these corresponding brain regions are strongly active during task execution (i.e., while participants viewed the videos in the dataset). This elevated power suggests increased engagement of these cortical areas under Alpha-dominant conditions. These regions are therefore associated with relaxed wakefulness or lower cognitive load. In contrast, Fig. \ref{Topo_Alpha_Gamma} (b) shows the EEG topographic map of the Gamma band, which reveals elevated power in regions corresponding to the T8, FPZ, FP1, FP2, PO5, AF4, and AF3 channels. This pattern indicates increased neural engagement in these areas during Gamma-dominant states. Higher power intensity indicates a stronger contribution of the corresponding brain region to deep cognitive focus. Table \ref{comparison} depicts the clear comparison of the top 10 channels as well as their corresponding brain region, cognitive functionality, and patterns in the Alpha and Gamma bands.


\begin{figure} 
    \centering
  \subfloat[\label{}]{%
       \includegraphics[width=0.5\linewidth, trim=0 102 60 120,clip]{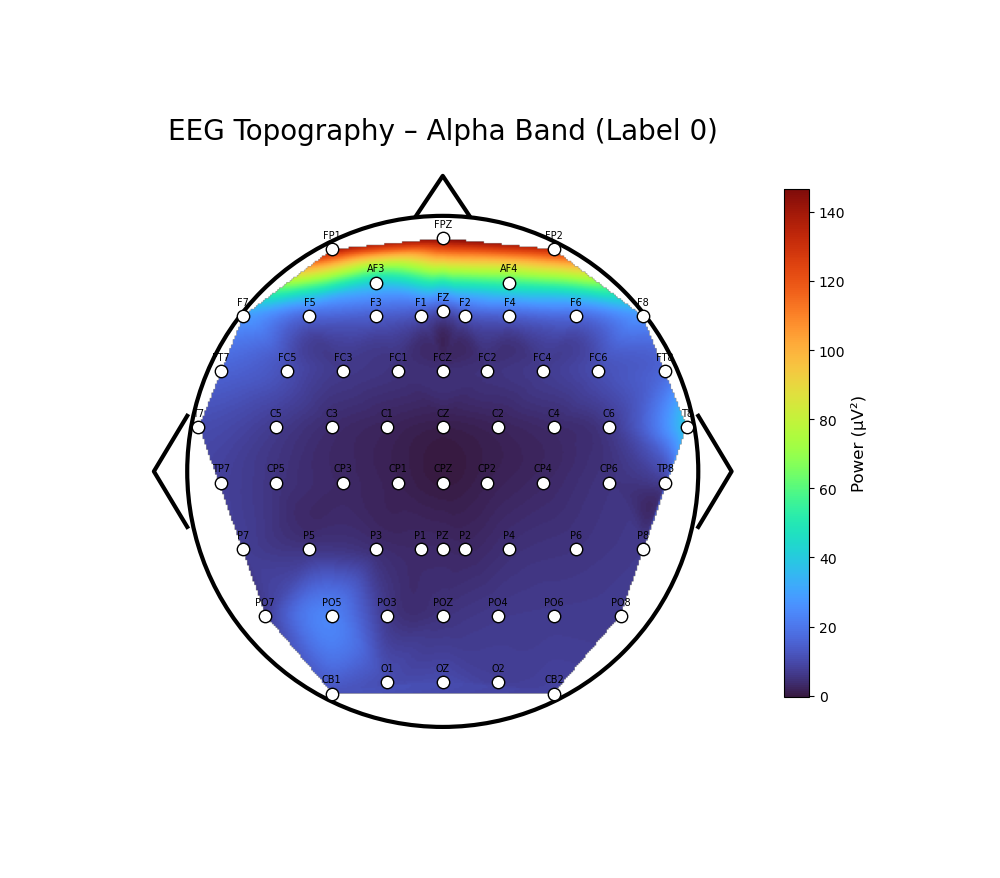}}
    \hfill
  \subfloat[\label{}]{%
        \includegraphics[width=0.5\linewidth, trim=0 102 60 120,clip]{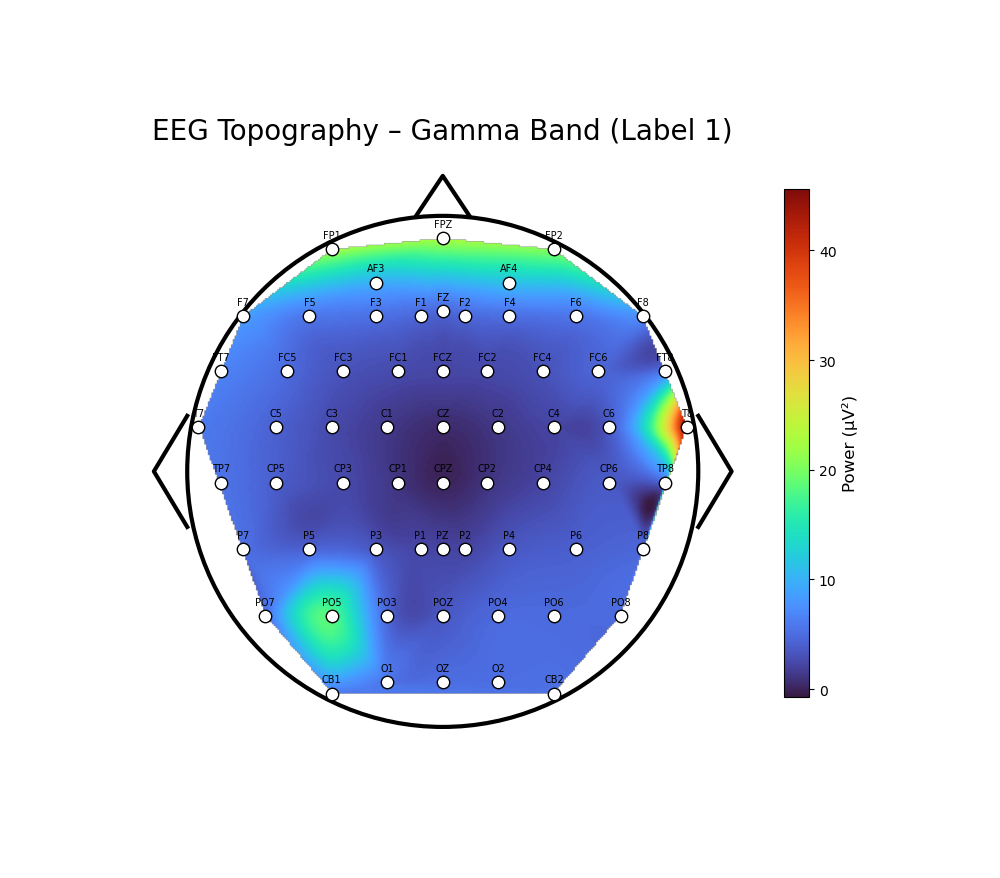}}
    \\

  \caption{EEG topographical map for (a) Alpha band and (b) Gamma band.}
  \label{Topo_Alpha_Gamma} 
\end{figure}

\subsubsection{\textit{Eye Tracking Topographical Map}} Fig. \ref{eye_alpha_gamma_topo} (a) shows the topographical map for eye tracking Alpha band data, which exhibits patterns consistent with those shown in Fig. \ref{Eye_Alpha}. In the Alpha band, higher power is indicative of a low attention, low cognitive demand state. In contrast, Fig. \ref{eye_alpha_gamma_topo} (b) illustrates the topographical maps derived from fixation, saccade, and pupil features for the Gamma band, showing patterns similar to those observed in Fig. \ref{Eye_Gamma}. Furthermore, Table \ref{eye}, highlights the relevance of eye-tracking measures in relation to corresponding EEG channels.



\begin{figure} 
    \centering
  \subfloat[\label{}]{%
       \includegraphics[width=0.5\textwidth, trim=0 0 20 30,clip]{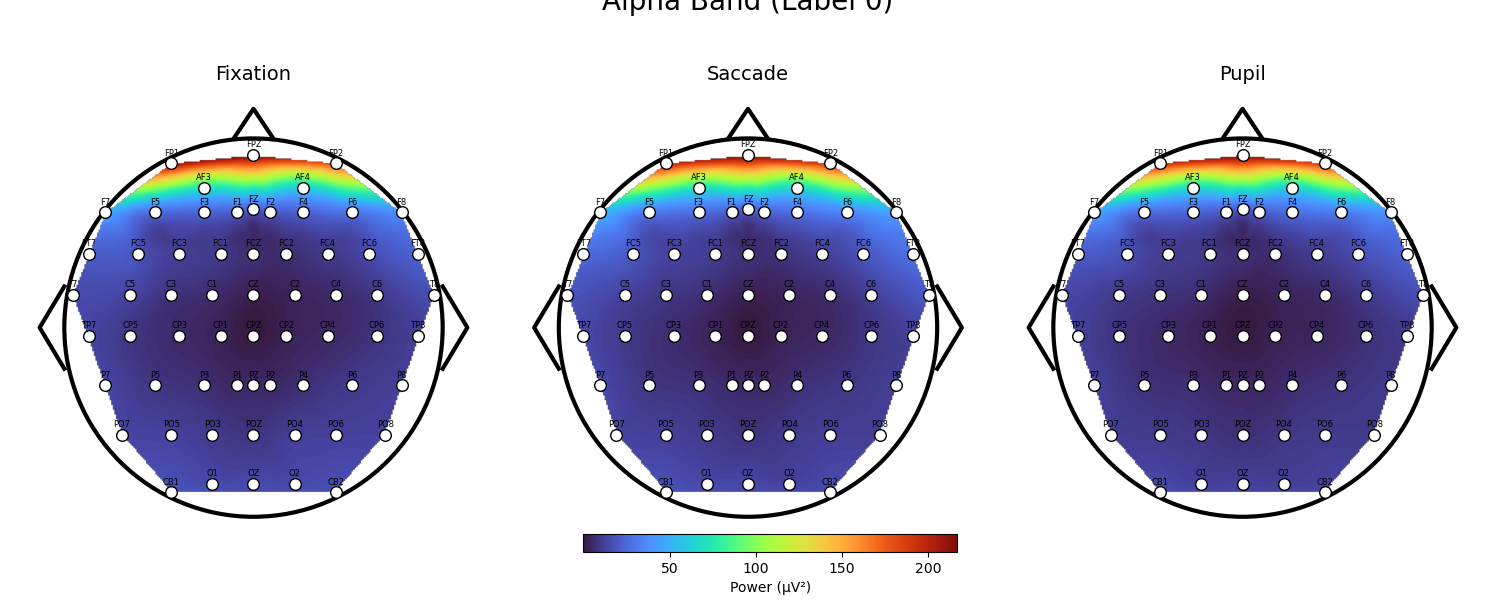}}
    \hfill
  \subfloat[\label{}]{%
        \includegraphics[width=0.5\textwidth, trim=0 0 0 20,clip]{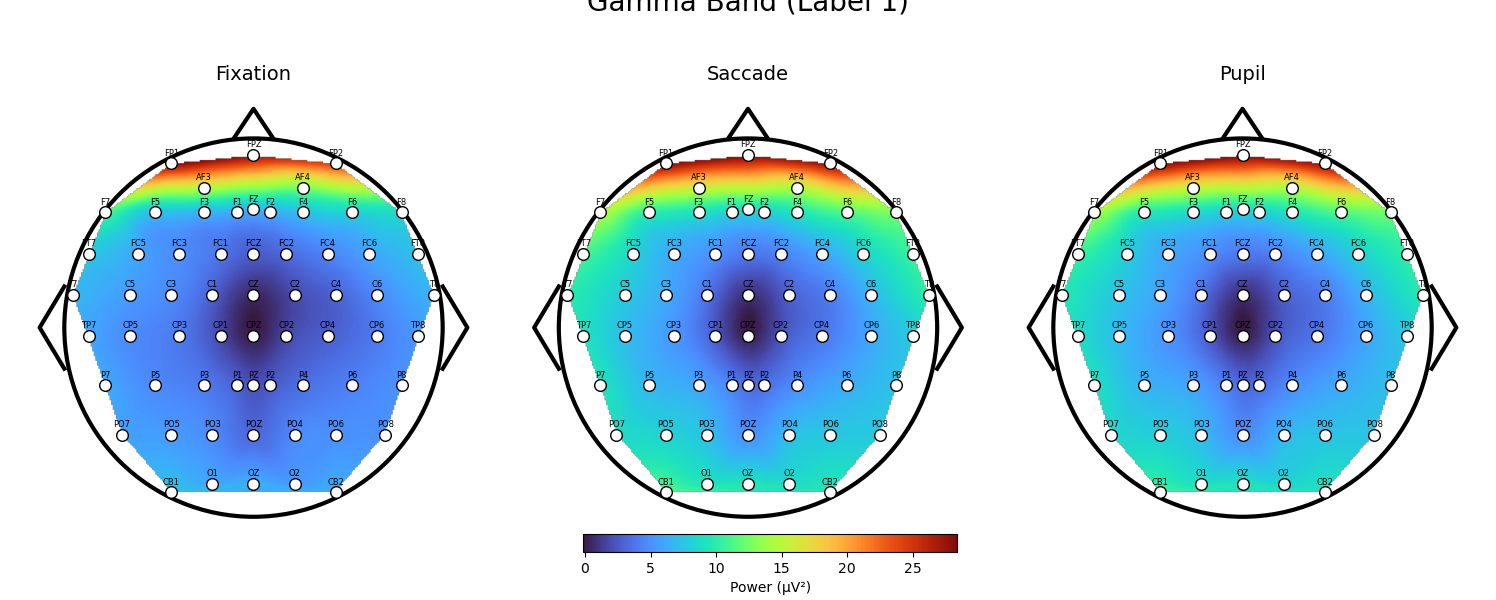}}
    \\

  \caption{Topographical map for eye tracking (a) Alpha band and (b)  Gamma band.}
  \label{eye_alpha_gamma_topo} 
\end{figure}



\subsection{Discussion}

In this study, the brain regions that are most influential for the deep focus state are extracted. From the EEG Gamma band data, it is found that T8 (Temporal
Right), FPZ (Frontopolar Midline), FP1 (Frontopolar Left),
FP2 (Frontopolar Right), PO5 (Parieto-Occipital Left), AF4
(Anterior Frontal Right), AF3 (Anterior Frontal Left), F7
(Frontal Left Lateral), F8 (Frontal Right Lateral), and FT7
(Frontotemporal Left) are the channels that contain the highest signal power. As the gamma band is responsible for the deep focus state, it is clear that Temporal, Frontopolar, Parieto-Occipital, Frontal, Frontotemporal, and Anterior Frontal are the brain regions that are responsible for high-level cognition or deep focus. To verify the findings derived from the EEG analysis, eye-tracking data are also utilized in this study.  Analysis of the eye-tracking data indicates that the Frontopolar, Frontal, Frontotemporal, and Anterior Frontal regions are the most influential. Overall, both EEG and eye-tracking analyses identify largely overlapping cortical regions as important for deep cognitive attention.
One notable exception is the T8 (Temporal Right) region, which emerges as highly influential in the EEG analysis but does not show comparable importance in the eye-tracking results. The T8 channel is prominent in the EEG analysis because it reflects deep cognitive processing and strong Gamma-band activity, particularly during tasks with high cognitive load. In contrast, T8 does not emerge as influential in the eye-tracking analysis, as eye movements are primarily governed by frontal and ocular motor regions rather than the temporal cortex where T8 is located. 
As shown previously in Fig. \ref{alpha_gamma_LIME}
and \ref{eye_gamma_lime}, the channels that influence the classification of Gamma- and Alpha-dominant activity can be identified. Careful examination reveals that the brain regions highlighted by the power intensity analysis are largely consistent with those identified through LIME-based explainability.  
 
Overall, the Alpha and Gamma bands exhibit largely overlapping sets of channels and brain regions with higher signal strength. A key distinction between the Alpha and Gamma topographical maps lies in their overall power strength. Specifically, the Alpha band exhibits higher overall power strength compared to the Gamma band. Notably, the Alpha power intensity of a few channels, such as FPZ, FP1, FP2, AF3, and AF4, is relatively higher than the corresponding Gamma-band power in these regions. Channels, such as T8, C1, CZ, C2, and CPZ, exhibit similar power levels in both the Alpha and Gamma bands (see Fig. \ref{alpha_gamma_power} for Alpha-Gamma power comparison). Besides, the Gamma band's overall topographic map is brighter than the Alpha band's map. In the Gamma band, brighter colors indicate higher power intensity. This suggests that most brain regions (except the central area) are engaged during deep cognitive focus (see Fig. \ref{eye_alpha_gamma_topo} (b)). 

This work further analyzes high-frequency neural dynamics by extracting Gamma-band power and burst-related features, including burst rate and duration, from the EEG signals. The Gamma-band power demonstrates substantial spatial variability, with a maximum of 46.69 µV² observed at channel T8 and a minimum of 0.089 µV² at CPZ. A consistent increase in burst rate, burst count, and burst duration is observed in the Gamma band compared to the Alpha band. Such elevated burst activity is indicative of increased cognitive load and sustained attentional engagement, as Gamma oscillations are commonly associated with high-level cognitive processing. Furthermore, longer burst durations may reflect enhanced temporal synchronization within cortical networks, suggesting more stable and coordinated neural activity during deep cognitive focus.


\section{Conclusion}
 Deep cognitive focus is associated with heightened Gamma-band oscillations and coordinated visual behavior. This study introduces Gamma2Patterns, a multimodal framework that examines EEG power and burst dynamics in the Gamma and Alpha bands alongside eye-tracking measures. Using the SEED-IV dataset, we identify key brain regions, including the frontopolar, frontal, frontotemporal, and anterior frontal areas that consistently exhibit strong Gamma power, elevated burst activity, and corresponding eye-movement patterns. Together, these findings highlight the critical role of these regions in sustained and profound attentional engagement. Moreover, we demonstrate that Gamma power and burst duration serve as more distinctive indicators of deep focus than Alpha power alone, highlighting their significance for attention decoding. In addition, classification and explainability analyses revealed that power-based features are sufficient to reliably distinguish Alpha- and Gamma-dominant states, achieving up to 90\% accuracy for EEG-based classification and comparable performance using eye-tracking features. While these predictive results validate feature relevance, the primary contribution of this study lies in its physiologically grounded mapping of deep-focus–related brain regions, moving beyond classification toward interpretable neural characterization. Overall, this work provides a multimodal, evidence-based map of brain regions and oscillatory patterns underlying deep cognitive attention, offering a neurophysiological foundation for future research in attention modeling and brain-inspired intelligence.
\section{Acknowledgment}
This work was supported by the BRAINS Lab (Bioinspired Robotics, AI, Imaging \& Neurocognitive Systems Laboratory) in the Department of Computer Science at The University of Alabama. 
\bibliographystyle{IEEEtran}
\bibliography{ref}
\end{document}